# Coaxial GaAs/(In,Ga)As Dot-in-a-Well Nanowire Heterostructures for Electrically-Driven Infrared Light Generation on Si in the Telecommunication O Band

*Jesús Herranz,\* Pierre Corfdir,[a] Esperanza Luna, Uwe Jahn, Ryan B. Lewis,[b] Lutz Schrottke, Jonas Lähnemann, Abbes Tahraoui, Achim Trampert, Oliver Brandt, and Lutz Geelhaar*

Paul-Drude-Institut für Festkörperelektronik, Leibniz-Institut im Forschungsverbund Berlin e.V., Hausvogteiplatz 5—–7, 10117 Berlin, Germany

[a] Present address: ABB Corporate Research, 5405 Baden-Dättwil, Switzerland

[b] Present address: Department of Engineering Physics, McMaster University, L8S 4L7 Hamilton, Canada

\*Email: herranz@pdi-berlin.de

## Abstract

Core-shell GaAs-based nanowires monolithically integrated on Si constitute a promising class of nanostructures that could enable light emitters for fast inter- and intrachip optical connections. We introduce and fabricate a coaxial GaAs/(In,Ga)As dot-in-a-well nanowire heterostructure to reach spontaneous emission in the Si transparent region, which is crucial for applications in Si photonics. Specifically, we achieve room temperature emission at 1.27 µm in the telecommunication O band. The presence of quantum dots in the heterostructure is evidenced by a structural analysis based on scanning transmission electron microscopy. The spontaneous emission of these nanowire structures is investigated by cathodoluminescence and photoluminescence spectroscopy. Thermal redistribution of charge carriers to larger quantum dots explains the long wavelength emission achieved at room temperature. Finally, in order to demonstrate the feasibility of the presented nanowire heterostructures as electrically driven light emitters monolithically integrated on Si, a light emitting diode is fabricated exhibiting room-temperature electroluminescence at 1.26 µm.

## Introduction

InAs quantum dots (QDs) embedded in strain-reducing (In,Ga)As layers, so-called dot-in-a-well (DWELL) structures, have been established as suitable heterostructures to efficiently extend the emission range of the InAs/GaAs material system to the 1.3 µm spectral region.[1,2] In particular, these structures have found application as the active region of optoelectronics devices monolithically integrated on Si.[3,4] However, the heteroepitaxy of III-As semiconductors on Si substrates for such applications is challenging and requires thick buffer layers between the Si substrate and the active region. Furthermore, advanced designs, such as the addition of layers acting as dislocation filter to reduce the density of threading dislocations and minimize their impact on the device performance, have to be employed. In the context of III-V integration on Si substrates, nanowires (NWs) have been recognized as a promising alternative approach.[5] The ability of NWs to accommodate large lattice-mismatches without introducing extended defects, thanks to the high surface-to-volume-ratio and small footprint on the substrate,[6] and the possibility to realize heterostructures in a core-shell geometry,[7] which provides a large active area in relation to the footprint of the NWs on the substrate, make these structures excellent candidates for the development of optolectronic devices such as laser or light emitting diodes (LEDs).[8–10]

In this work, we demonstrate the growth of a coaxial GaAs/(In,Ga)As DWELL NW heterostructure, combining the NW geometry for integration on Si substrates with the DWELL heterostructure that enables to shift the emission to longer wavelengths. In fact, this heterostructure allows us to reach the transparent window of Si and, in particular, achieve room temperature operation in the telecommunication O band (1.26–1.36 μm), demonstrating a new functionality of a heterostructure integrating QDs with NWs. Hierarchical structures combining QDs and NW are a subject of active research, both in axial and radial geometries.[11] In radial heterostructures, the main material system of study that has attracted strong interest is constituted by (Al,Ga)As shells with spontaneously forming local enrichments in Ga giving rise to QD states.[12–16] These QDs results from alloy fluctuations and their optical emission is typically in the range 1.8-1.9 eV. In general, structures combining QDs and NW have so far been considered for specific applications such as single photon emitters[12,17] or opto-mechanical systems for sensing,[18] but not to extend the spectral range of light emission. To realize this novel structure, we had to overcome the fact that InAs QDs do not form on the {110} sidewall facets of GaAs NWs.[19] Thus, we exploited the recently demonstrated capability of Bi surfactants to promote QD formation on GaAs NW sidewalls.[20,21] Normally, if InAs is deposited on GaAs{110} surfaces, including the {110} sidewall facets of self-seeded GaAs NWs, strain relaxes plastically before the Stranski-Krastanov transition from two- to three-dimensional growth takes place.[22] The use of a Bi surfactant modifies surface energies, enabling the formation of InAs QDs to develop a coaxial DWELL structure.

In addition to a structural analysis based on scanning transmission electron microscopy (STEM), we present characterization of the optical emission properties of the DWELL NWs by cathodoluminescence (CL) and photoluminescence (PL) spectroscopy. Furthermore, in order to establish the feasibility of the presented heterostructures as electrically driven light emitters monolithically integrated on Si, we demonstrate room temperature operation of a DWELL NW LED with emission wavelength of 1.26 μm.

**Results and discussion**

All samples studied in this work were grown by molecular beam epitaxy (MBE) on Si(111) substrates. The main steps of the MBE growth are schematically presented in the upper part of Figure 1: Ga is first deposited to form Ga droplets (Figure 1a), which induce the axial growth GaAs NWs of about 100 nm diameter by the vapor-liquid-solid (VLS) method (Figure 1b). After consumption of the Ga droplet, the coaxial DWELL heterostructure is radially grown on the NW facets (Figure 1c). This multi-shell heterostructure consists of an InAs QD shell embedded in an about 9 nm-thick $In_{0.15}Ga_{0.85}As$ quantum well (QW) shell. A Bi flux was supplied during the deposition of InAs to promote QD formation.[20,21] A 50 nm thick outer GaAs shell completes the confinement in the DWELL heterostructure (Figure 1d). The total diameter of the DWELL NWs is approximately 220 nm, and a cross-sectional schematic of the coaxial heterostructure is depicted as an inset in Figure 1e. An extended description of the growth procedure is provided in the Experimental section. Scanning electron micrographs of a DWELL NW sample synthesized on a patterned Si(111) substrate are shown in Figure 1e. A higher magnification micrograph is included as inset.

Side-view scanning electron micrographs of uncapped InAs layers deposited with and without Bi supply on GaAs NWs with 2 nm thick $In_{0.15}Ga_{0.85}As$ shell layers are presented in Figure 1f and Figure 1g, respectively, in order to illustrate the effect of Bi during the InAs deposition. In the absence of Bi, a discontinuous plate-like shell structure is observed (Figure 1f), while small island-like features can be identified when Bi is simultaneously supplied to the InAs deposition (Figure 1g). These results are consistent with previous reports for growth directly on GaAs NW facets[19,21] and demonstrate that the effect of Bi on InAs deposition is analogous for the NW sidefacets with thin $In_{0.15}Ga_{0.85}As$ shells.

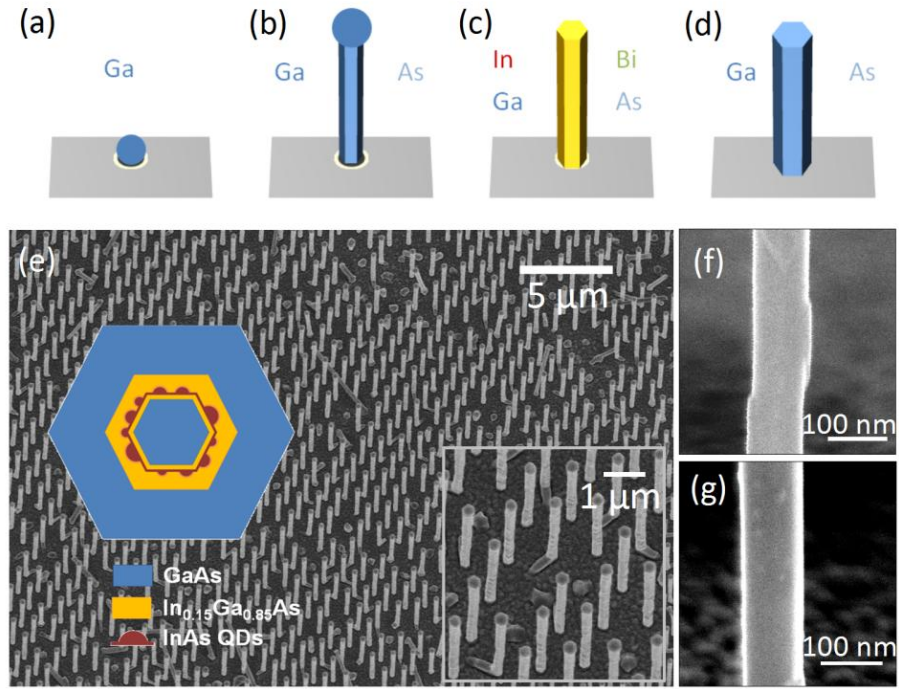

**Figure 1.** (a-d)Schematics of the main growth steps of the DWELL NWs: (a) Ga predeposition, (b) VLS GaAs NW core growth, (c) coaxial (In,Ga)As DWELL shell, and (d) GaAs outer shell growth. Labels indicate the fluxes used in each stage. (e) Scanning electron micrograph of DWELL NWs grown on a patterned Si(111) substrate. The pattern was fabricated using electron beam lithography and comprises a hexagonal array of holes with a diameter of 50 nm and a pitch of 1 µm. The diameter of the DWELL NWs is approximately 220 nm and the length is about 2.5 µm. A higher magnification micrograph and a cross-sectional schematic of the DWELL NW heterostructure are included as insets. (f,g) Side-view scanning electron micrographs of 2ML InAs layers deposited on about 100 nm diameter GaAs NWs with 2 nm thick $In_{0.15}Ga_{0.85}As$ shells (f) without simultaneous Bi supply and (g) with simultaneous Bi supply (g).

Microstructural characterization by STEM of a DWELL NW is presented in Figure 2. A cross-sectional micrograph acquired using the scanning high-angle annular dark field (HAADF) mode[23] is shown in Figure 2a. In this imaging mode the contrast is mainly determined by the atomic number $Z$ (Z-contrast). The NW exhibits a hexagonal cross-section defined by {110} side facets and the coaxial active region is clearly visible in the HAADF micrograph. An enlarged false-color micrograph of the facet indicated by the dashed box in Figure 2a is presented in Figure 2b to illustrate the presence of QDs in the DWELL heterostructure. The yellow areas indicate higher HAADF intensity, corresponding to a locally enhanced In composition within the projected volume, which can be attributed to the QDs.

A quantitative analysis is presented in the lower part of Figure 2. HAADF intensity ($I_{HAADF}$) profiles taken for two different facets highlighted in Figure 2a are depicted in Figures 2c and 2d, respectively. In Figure 2c (blue profile), the $I_{HAADF}$ level for the DWELL active region is almost constant, which we attribute to an (In,Ga)As QW segment where QDs are not present/detected. The probability of finding a QD in the thin (<100 nm) electro-transparent specimens is determined by the random distribution of the QDs along the NW. In contrast, in Figure 2d (magenta profile) the $I_{HAADF}$ profile shows higher intensity spots (indicated by black arrows) within the DWELL active region, that correspond to the extended yellow areas in Figure 2b. As introduced above, we attribute these local variations in $I_{HAADF}$ to InAs QDs present in the DWELL heterostructure. The slope in the $I_{HAADF}$ profile in Figure 2d arises from a small gradient in the thickness/tilt of the TEM NW-specimen. The In content profile obtained for this facet of interest is presented in Figure 2e. The estimated In distribution is obtained from the analysis of the noise-corrected HAADF intensity profile (corresponding to the area highlighted in Figure 2c) following the procedure reported in Ref. [18]. In brief, the method relies on the dependence of the HAADF contrast $R = I_{InGaAs(x)}/I_{GaAs}$ on the In content $x$[24,25] after correcting for specimen thickness variations and assuming $\gamma = 1.7$ in $I_{HAADF} \propto Z^{\gamma}$. The suitability of $\gamma = 1.7$ is determined from measurements on planar reference samples of known composition. The estimated In content in the QDs lies in the range $x = 17–27\%$. Similar data for two additional DWELL NWs is included in Supporting Information, with a QD In content in the range $x = 18–25\%$. We should note that the In content that we estimate is averaged over the specimen thickness and that the reported values should be thus taken as lower bounds. Hence, the true In content associated to the QDs is probably higher. Furthermore, for the same reason the actual shape of the QDs and details like possible intermixing cannot be deduced from this data. However, studies of Bi-induced QDs grown on GaAs(110) planar surfaces and NW sidewalls suggest the presence of a wetting layer and a QD shape with a much larger width than height.[21,26,27] The determination of the actual In content and distribution within the QDs introduced here, as well as their precise shape, would require additional advanced microstructural characterization beyond the scope of this work.[19,28]

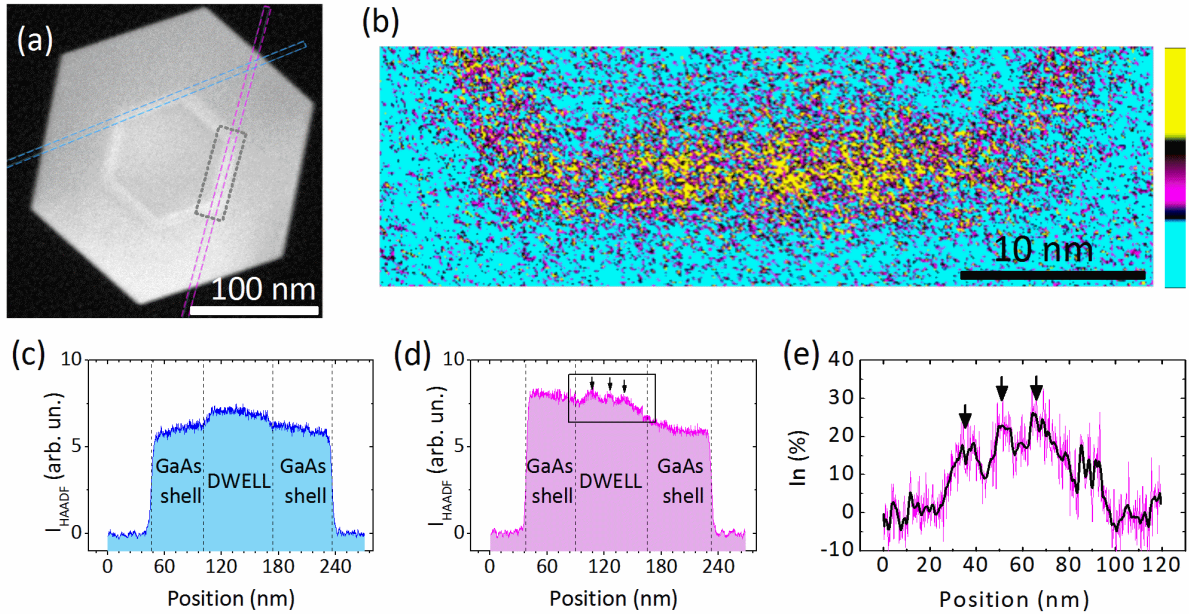

**Figure 2.** (a) Plan-view HAADF STEM micrograph of a DWELL NW. (b) Enlarged, false-color, baseline corrected HAADF STEM micrograph of the facet indicated by the dashed box in (a). (c,d) HAADF intensity $I_{HAADF}$ profiles taken along the blue and magenta boxes sketched in (a), respectively. (e) Indium profile extracted from the noise-corrected $I_{HAADF}$ profile corresponding to the area highlighted in (d) by the black box. The raw data (magenta line) are presented along with smoothed data (black line) for clarity. The black arrows indicate local In content increments, consistent with the presence of (In,Ga)As QDs in the DWELL heterostructure.

The luminescence of the DWELL NWs is first studied by CL spectroscopy as presented in the upper part of Figure 3 (Figure 3a-d). A low temperature CL spectrum is displayed in Figure 3a. The spectrum corresponds to the averaged emission of the area presented in Figures 3b and 3c, containing about 30 NWs, and is dominated by a broad band centered at 1.1 eV (1.13 µm). At higher energy, there is a second band of lower intensity peaking at 1.35 eV (0.92 µm). We attribute this high energy band to transitions in the (In,Ga)As QW and the low energy band to transitions in (In,Ga)As QDs embedded in the QW. In similar NWs with only an (In,Ga)As shell QW that was grown without supply of Bi and without InAs deposition, only the high energy band was observed.[29–31] Likewise, the Supporting Information contains the low temperature photoluminescence (PL) spectrum of a dedicated reference sample that was grown under the same conditions as the sample analyzed in Figure 3 but without Bi supply and without insertion of an InAs layer. This spectrum does not contain the low energy band. Moreover, the low energy band is not observed, either, for another reference sample that was grown with InAs insertion but without Bi supply. Further arguments for assigning the low energy band to QDs are discussed in the following. The broad energy distribution of the low energy band seen in Figure 3a reflects the variations of dimensions and composition of the QDs. Monochromatic CL maps collected at the peak energies 1.345 eV and 1.088 eV are presented in Figure 3b and Figure 3c, respectively. The CL map in Figure 3b shows that the high energy emission originates from an approximately 350 nm long segment close to the top of the NWs, that we assign to a wurtzite (WZ) (In,Ga)As coaxial QW segment, as we have recently reported for similar structures without QDs.[31] The formation of a WZ segment on top of zincblende (ZB) GaAs NWs during core growth is related to the consumption of the Ga droplets prior to shell growth. The CL map in Figure 3c shows that the low energy emission originates from spatially localized states distributed along the NWs, consistent with its attribution to QD states. A more detailed analysis of the CL data reveals that the QW emission from the ZB region is very weak, pointing to an efficient carrier transfer to the QDs. Interestingly, we observe no overlap between the WZ QW segment and the areas with localized QD emission. This suggests that QDs do not form in the WZ segments of the NWs. In Figures 3b and 3c, no CL signal is detected from the parasitic layers grown on the substrate, as expected for the polycrystalline material deposited on the amorphous oxide layer.

An additional hyperspectral CL linescan acquired for a single DWELL NW is presented in Figure 3d, which further corroborates the attribution of the low energy emission band to QDs. Spatially localized sharp spectral features assigned to the recombination in the QDs are present along the NW. These emission lines are elongated along the NW axis, which can be explained by diffusion of carriers excited by the electron beam to the QDs. It is important to note that the limited spectral range of the Si detector used to record the linescan (cut-off at ≈ 1060 nm) does not allow us to investigate the broad low energy emission band of the DWELL NWs; and the QDs revealed in this CL linescan only correspond to the smaller ones in the NW.

The bottom part of Figure 3 presents low temperature PL spectra of as-grown DWELL NWs. In these experiments, only 2-3 NWs are excited simultaneously. Sharp spectral features are revealed at low excitation power density (Figure 3e), similar to the transitions observed for InAs QDs on GaAs NW facets.[26] Figure 3f shows in linear scale the spectral range marked in Figure 3e. The spectral linewidth of the most prominent transitions is approximately 200 µeV, limited by the spectral resolution of the experimental setup. The sub-meV spectral linewidth shows that these transitions originate from localized excitons, consistent with optically active QDs in the DWELL heterostructure.

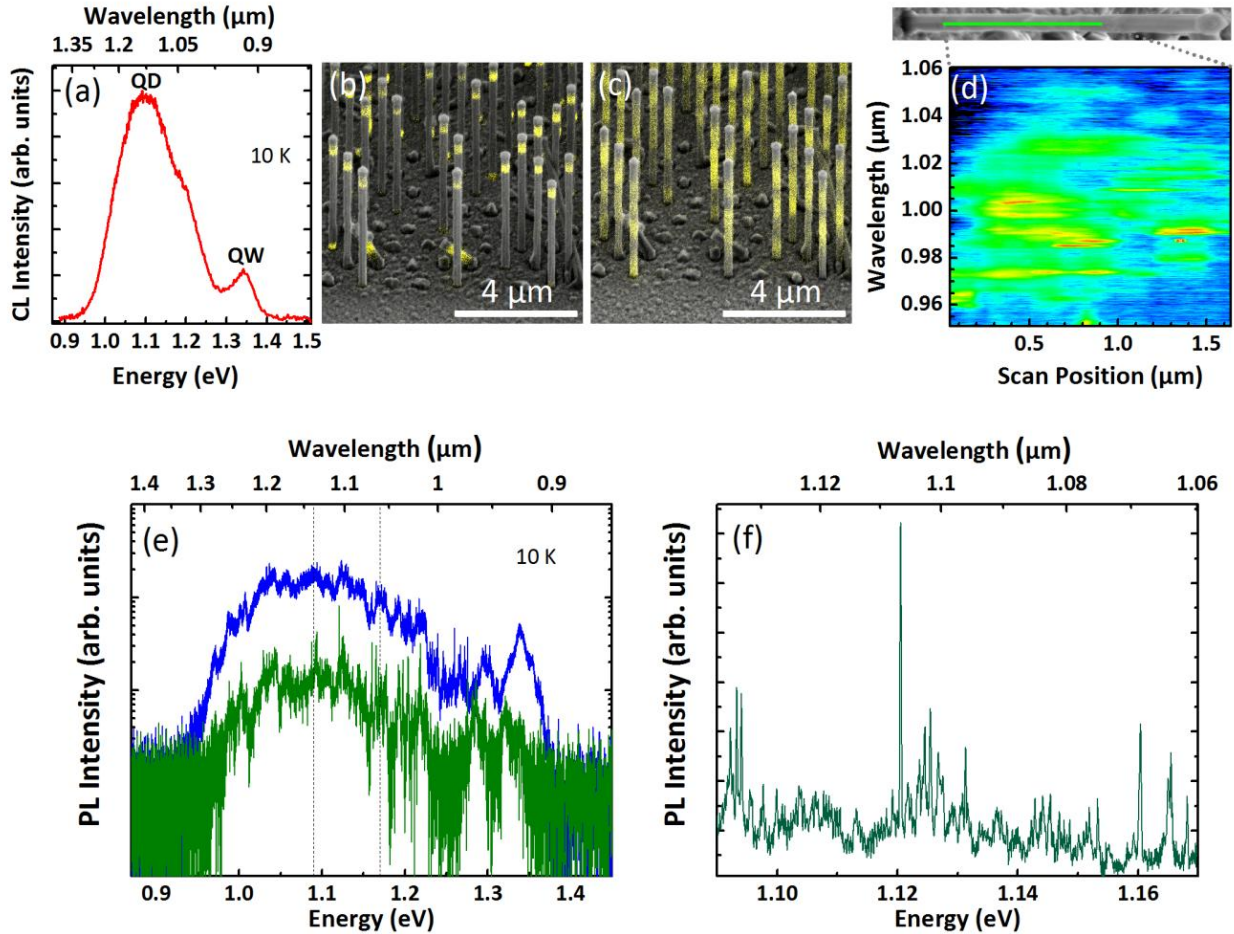

**Figure 3.** (a) Low temperature (10 K) CL spectrum of an as-grown DWELL NW sample. (b,c) Monochromatic, false-color CL maps acquired at (b) 1.345 eV (0.922 µm) and (c) 1.088 eV (1.140 µm) superimposed on the corresponding birds-eye-view scanning electron micrograph of the NW sample. (d) 2D plot of the CL emission acquired in a linescan along the axis of a single DWELL NW. The spectral range is limited by the response of the Si CCD detector. A scanning electron micrograph of the studied NW is displayed above the CL map. The green line indicates the linescan. (e) PL spectra of few (2–3) as-grown DWELL NWs acquired at low temperature (10 K). (f) PL spectrum of DWELL NWs displaying sub-meV transitions in the spectral range indicated in (e) revealing QD states in the DWELL heterostructure. The spectral linewidth of the most prominent transitions observed in (f) is approximately 200 µeV, limited by the spectral resolution of the experimental setup.

In addition, we investigate the spontaneous emission of the DWELL NWs by PL spectroscopy as a function of temperature from 10 to 300 K in Figure 4. In these experiments, around 60–70 NWs are excited simultaneously. The temperature-dependent PL spectra are presented in Figure 4a. Again, we observe the two different emission bands seen in the CL spectrum (Figure 3a), with a high energy component (1.35 eV at 10 K) attributed to QW transitions and a low energy component (1.10 eV at 10 K) attributed to QD transitions. At room temperature, the emission is dominated by the low energy QD component.

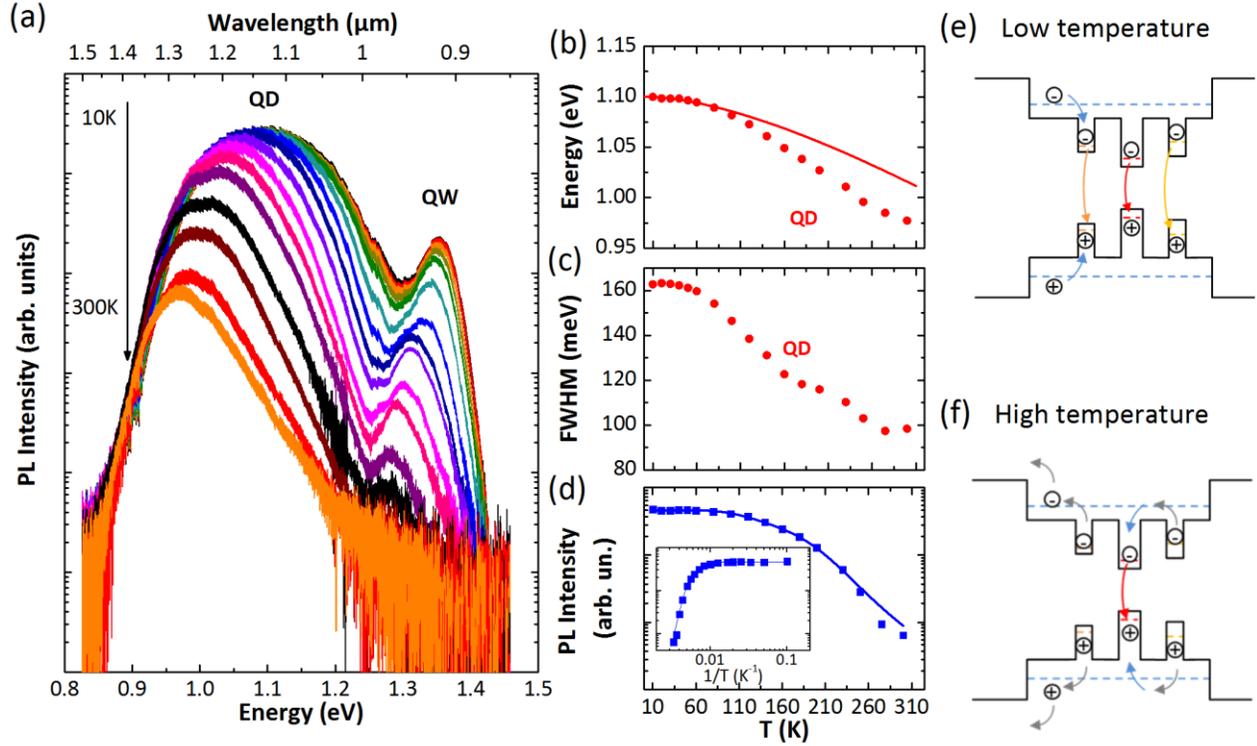

**Figure 4** (a) PL spectra of a DWELL NW sample measured from 10 to 300 K (excitation of around 60–70 NWs). Low and high energy bands are labeled as QD and QW as discussed in the main text. (b) Peak emission energy and (c) FWHM of the low energy QD band. The red solid line in (b) corresponds to the expected temperature dependence of a bulk $In_{0.385}Ga_{0.615}As$ random alloy and it is included for comparison to highlight the redshift in the DWELL NW structure. (d) Integrated intensity $I_{PL}$ of the DWELL NWs as a function of temperature. The black solid curve in (d) is a fit of the experimental data to the phenomenological model described in the main text with the activation energies $E_1 = (44 \pm 5)$ meV and $E_2 = (188 \pm 37)$ meV. An Arrhenius representation of the same data is included as an inset. (e, f) Schematics of the main recombination processes at low (e) and high (f) temperature. These schematic band diagrams include the QW energy level (blue line) and three different energy levels for QDs representing variations of size and/or In content. Low temperature emission takes place from the QW and essentially all QD levels, while high temperature emission is dominated by QDs with the lowest energy.

The evolution with temperature of the peak emission energy and full width at half maximum (FWHM) for the QD band are presented in Figures 4b and 4c, respectively. With increasing temperature, the emission energy shows a strong redshift to 0.97 eV at 300 K. We should note that the observed redshift is about 20 meV larger than that expected for a random $In_{0.385}Ga_{0.615}As$ bulk alloy with the same peak emission at low temperature,[32] as presented by the solid line in Figure 4b. Simultaneously, the FWHM is reduced from 162 meV at 10 K to 104 meV at 300 K. At room temperature, the peak emission wavelength of these DWELL NWs is 1.27 μm, thus reaching the O band (1.26–1.36 μm) used for data communication.

Figure 4d shows that the PL intensity $I_{PL}$ integrated over the full spectral range decreases with increasing temperature by about two orders of magnitude. We analyze this variation with a simple phenomenological model with two non-radiative channels for charge carriers: $I_{PL} = \frac{I_0}{1+a_1 exp\left(\frac{-E_1}{kT}\right)+a_2 exp\left(\frac{-E_2}{kT}\right)}$. $T$ is the temperature, $k$ is the Boltzmann constant, and $I_0$ is a constant. The non-radiative channels are described by the prefactors $a_1$ and $a_2$ and activation energies $E_1$ and $E_2$, respectively. Similar phenomenological models are used to describe classical planar DWELL structures.[33–35] A fit of the experimental data results in the activation energies $E_1$ = 44 ± 5 meV and $E_2$ = 188 ± 37 meV. These activation energies should be considered as an average description over all NWs in the ensemble. Interestingly, these values are similar to the activation energies ($E_1$ = 34 meV and $E_2$ = 262 meV) reported for a planar InAs/(In,Ga)As DWELL structure with (001) orientation.[35] Furthermore, the activation energy of the first channel $E_1$ = 44 ± 5 meV is similar to one of the decay channels reported for (In,Ga)As/GaAs NWs with shell QW[30] activation energy of the second channel $E_2$ = 188 ± 37 meV is comparable to the energy difference between the QW and QD bands observed in the low temperature PL spectra (≈ 250 meV) and, therefore, points to the thermal escape of carriers from the QDs to the QW. This finding along with the narrowing and the strong redshift of the emissionas well as the fact that the thermal quenching is moderate compared to four orders of magnitude intensity reduction observed for standard (In,Ga)As/GaAs shell QW NWs without any QDs,[30] suggest a thermal redistribution of charge carriers in the QDs of the DWELL heterostructure.

This mechanism is sketched in Figures 4e and 4f and it is well established for planar DWELL heterostructures.[35] At low temperature (Figure 4e), the charge carriers populate QD levels, as well as the QW levels (predominantly in the WZ segment in our case, due to the efficient carrier transfer in the ZB NW) and recombine mostly radiatively, thus contributing to the broad PL spectrum. With increasing thermal energy (Figure 4f), charge carriers can escape shallow energy levels of smaller QDs and undergo non-radiative recombination, as observed by the quenching of the PL intensity above 160 K, or are recaptured in larger QDs with deeper energy levels, which manifests itself in the narrowing and redshift of the emission. The deeper confinement in larger QDs provides protection from non-radiative recombination centres and explains the enhanced room temperature emission as well as the longer emission wavelength.

The presented optical characterization evidences that the emission range of the coaxial DWELL NWs lies in the Si transparent window. In order to demonstrate the feasibility of the presented DWELL NWs as electrically driven light emitters on a Si platform, we fabricated LEDs from self-assembled ensembles of DWELL NWs with a radial *p-i-n* junction. GaAs NW based LEDs on Si substrates operating in the near infrared region have already been demonstrated by different research groups.[8,9,36,37] An emission wavelength of 985 nm has been demonstrated for a coaxial multishell (In,Ga)As/GaAs NW LED,[37] and a nanoneedle LED with a similar (In,Ga)As/GaAs QW active region reached 1.13 μm,[38] which constitutes the longest wavelength of electroluminescence (EL) demonstrated for GaAs-based NW-like structures. For the present study, the LED multishell structure grown around an approximately 100 nm thick Be doped *p*-type GaAs NW core consists of a 10 nm thick undoped GaAs shell, a coaxial DWELL heterostructure, a 10 nm thick undoped GaAs shell, and an outer 50 nm thick *n*-type GaAs shell, which was doped with Si in an approach established previously.[39] The schematic cross section of the LED structure is shown as an inset in Figure 5b. The total diameter of the LED NWs is approximately 260 nm.

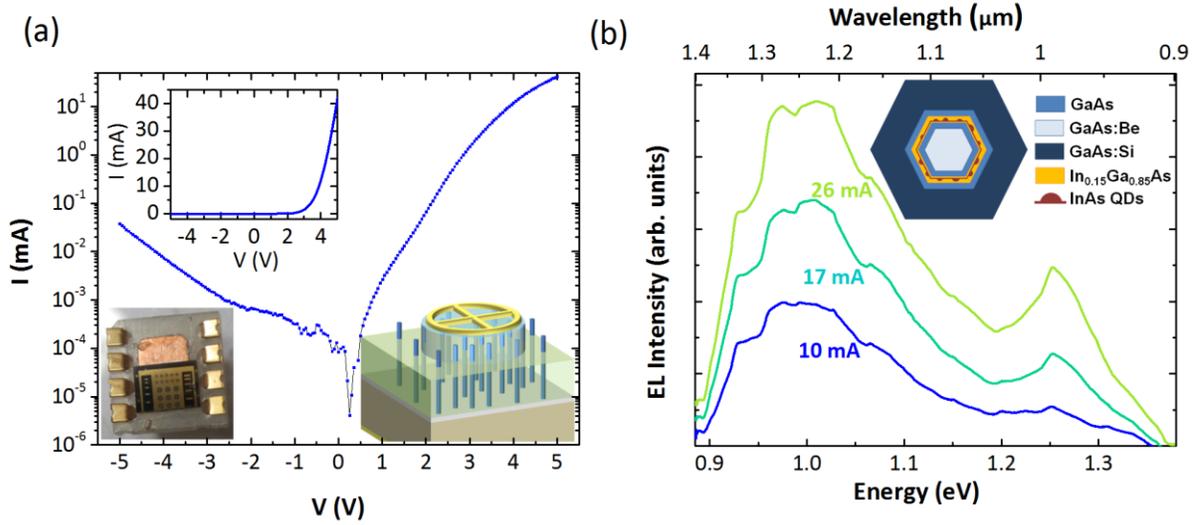

**Figure 5** (a) Current-voltage (*I-V*) characteristics of a DWELL NW LED with a diameter of 300 µm, acquired at room temperature. The same data are presented in linear scale in the inset. A schematic of a DWELL NW LED and a picture of the fabricated chip mounted in its chip carrier are included as further insets. (b) Room-temperature EL spectra of a DWELL NW LED driven at 10 mA, 17 mA and 26 mA. A cross-sectional schematic of the NWs with radial *p-i-n* junction is included as an inset.

The current-voltage (*I-V*) characteristics of a 300 µm diameter LED, contacting approximately $10^5$ NWs in parallel, are presented in Figure 5a, revealing rectifying behavior of the fabricated device. Nevertheless, we observe an exponential increase of the current at negative bias beyond −2 V in the reverse region of the I-V curve, which is probably caused by defect-supported intraband tunneling. The analysis of the forward region of the *I-V* curve using the standard diode equation $I = I_S \left[ exp\left(\frac{V-IR_S}{nkT}\right) - 1 \right]$ gives an ideality factor *n* = 15.2, a series resistance $R_S$ = 16.9 Ω and a saturation current $I_S$ = 8.40×10$^{-7}$ A. High ideality factors have been previously reported for core-shell NW LEDs[9] and, in addition to the turn-on voltage of about 2.5 V (larger than the GaAs bandgap), suggest that the actual device consists of more than one diode in series, as a result of additional barriers in the device. One additional barrier could be present at the interface between the Si substrate and the NWs. A limited incorporation of dopants in the initial stages of NW growth, as well as a possible indiffusion of Si atoms into the NW from the substrate, could result in a doping level too low to obtain a good contact to the *p*-type substrate. The thin native oxide layer and possible structural defects at the interface may also contribute. An additional barrier could be located in the top contact between the top transparent layer made of indium tin oxide (ITO) and the *n*-type GaAs outer shell of the NWs.[40] The study of the interfaces described above and optimization of the dimensions and doping profiles of the doped sections of the radial *p-i-n* junction in our devices are necessary to improve the electrical performance of the LEDs. In addition, the use of patterned substrates should lead to more homogeneous NW arrays and better isolation of the *p-i-n* junction from the substrate.

The EL spectra of one DWELL NW LED measured at room temperature at injected currents in the range 10–26 mA are presented in Figure 5b. The emission band peaks at 1.26 µm, consistent with the previously presented characterization of undoped structures. This result demonstrates the room temperature operation of electrically driven light emitters on Si in the telecommunication O band (1.26 –1.36 µm). Furthermore, it represents an extension of the emission wavelength by approximately 270 nm with similar brightness compared to our previous coaxial (In,Ga)As/GaAs QW NW LEDs[29] and by approximately 130 nm compared to GaAs based nanoneedle LEDs.[38] Optically excited, monolithically on Si integrated light sources based on GaAs NWs or related structures have been demonstrated up to 1.1 µm using GaAs-[(In,Ga)As/(Al,Ga)As] core–shell NWs with multiple (In,Ga)As QWs[41] and up to 1.44 µm using (In,Ga)As NWs.[42] More recently, the tunability of the GaAs bandgap up to 1.42 µm has been demonstrated exploiting strain partitioning in NWs.[43] LEDs operating in the telecommunication S band (1.46 –1.53 µm) have been demonstrated for InP based nanopillars on Si with radial (In,Ga)As shell QW.[44] Nanoneedle/nanopillar structures resemble NWs, but they typically exhibit base diameters in the micrometer range and are characterized by a tapered geometry, which has been demonstrated to sustain helicoidal guided photonic modes.[45] The DWELL NW heterostructure introduced here is grown around a GaAs NW core with diameter of about 100 nm and presents a smaller footprint on the substrate. This characteristic could be beneficial for the integration on photonic elements such as Si waveguides[46–48] and offers greater design flexibility to tune the total diameter by simply modifying the thicknesses of the GaAs shells, which would enable the tailoring of leaky or guided modes for specific applications.[49] Very exciting would be the use of DWELL NWs for lasers. The first step in this direction would be the formation of a cavity for optical confinement. In the simplest case, optical modes could be confined in a single NW of sufficient diameter and length. To provide high gain at the target wavelength, it would be desirable to narrow the emission spectrum of the QDs. We believe the corresponding optimization of the QD size distribution could possibly be achieved by modifying the way to induce QD formation with Bi. In particular, we recently reported that InAs QD formation can be triggered by exposing static InAs 2D layers to Bi, which induces a rearrangement of the strained layer into 3D islands on planar GaAs(110) surfaces.[27]

**Conclusion**

In conclusion, we have presented novel GaAs/(In,Ga)As DWELL NW heterostructures grown on Si substrates as sources for infrared light emission in the Si transparent spectral window. Microstructural characterization by STEM illustrates the presence of QDs in the DWELL heterostructure; and spectral widths below 1 meV, as observed in low temperature PL measurements, reveal the QD nature of the emission. The formation of QDs is deliberately triggered by supplying Bi during growth. Carrier redistribution to larger QDs in the structure by thermal escape and recapture explains the strong redshift of the emission with temperature and allows us to reach long wavelength emission at room temperature, similarly to planar DWELL structures. We have demonstrated room temperature operation of DWELL NW LEDs on Si at 1.26 µm; thus establishing the application of GaAs-based NWs for monolithically integrated electrically driven optoelectronics on a Si platform. These results illustrate the potential of heterostructures combining QDs and the NW geometry and push the electroluminescence from GaAs-based NWs into the spectral region relevant for integrated optical interconnectors and data communication, opening up new perspectives for Si photonics integration.

**Experimental section**

**Growth** – The samples were grown by solid-source MBE on Si(111) substrates. The MBE system is equipped with In, Al, Bi and two Ga effusion cells, Be and Si effusion cells for dopants and two valved cracker sources for supply of $As_2$. Fluxes are calibrated and expressed in the following in terms of equivalent growth rate on a GaAs (001) surface in monolayers per second (ML/s). An optical pyrometer was used for measuring the substrate temperature.

DWELL NWs were grown on *p*-type (B doped) Si(111) substrates covered with a 20 nm thick thermal oxide mask patterned by electron beam lithography (EBL). Before loading into the MBE system, patterned substrates were etched with a 1% HF solution and treated in boiling water for 10 minutes.[50] Ga was deposited on the substrate for 70 s at 0.3 ML/s GaAs equivalent growth rate to form Ga droplets on the surface and GaAs NWs were then grown by the self-catalyzed vapor-liquid-solid (VLS) method at 630 °C with a Ga flux of 0.3 ML/s and V/III ratio of 3.9 for 30 minutes. Then, the Ga flux was closed and the Ga droplets atop of the NWs were consumed by crystallization to GaAs under a 4 ML/s $As_2$ flux at the growth temperature of 630 °C. Subsequently, the DWELL structure, which consists of a 2 nm thick $In_{0.15}Ga_{0.85}As$ shell, a 0.3 nm (2 ML) thick InAs quantum dot (QD) shell and a 7 nm thick $In_{0.15}Ga_{0.85}As$ shell, was grown as active region around the GaAs NW core at 420 °C, a growth rate of 0.2 ML/s and a V/III flux ratio of 20. Growth parameters were chosen to limit adatom diffusion in order to obtain homogeneous shells.[37] Samples were rotated at 6 revolutions per minute during growth and, due to the inclination of the MBE sources (33.5°), NW sidewall growth rates are about 0.21 times the planar growth rate. A Bi flux at a beam equivalent pressure of $1.3 \times 10^{-6}$ mbar was supplied during InAs deposition in order to promote QD formation.[20,21] A 50 nm thick undoped GaAs shell grown under the same conditions completes the structure in order to provide carrier confinement in the DWELL region. An additional reference sample for which Bi was not supplied during InAs shell deposition was fabricated under nominally identical growth parameters. Similarly, a second reference sample for which neither Bi nor binary InAs was supplied, i.e. a NW sample with a $In_{0.15}Ga_{0.85}As$ coaxial QW active region, was grown with same growth parameters.

The DWELL NW sample for LED fabrication was grown on a 2 inch *p*-type (B doped) Si(111) wafer covered with its native oxide. Ga was deposited on the substrate for 30 s at 1.3 ML/s GaAs equivalent growth rate followed by a 60 s flux interruption to form Ga droplets on the surface. Be-doped GaAs NWs were grown at 645 °C with a Ga flux of 0.3 ML/s and V/III ratio of 2.4 for 30 minutes by the VLS method. Doping of the NWs was achieved by simultaneous supply of Be during VLS growth. The Be flux was adjusted to avoid morphological changes in the NWs.[51] Next, after consumption of the Ga droplets under a 4 ML/s $As_2$ flux, shells were grown around the NW core at 420 °C, a growth rate of 0.2 ML/s and a V/III flux ratio of 20. The active region of the DWELL heterostructure structure consists of a 2 nm thick $In_{0.15}Ga_{0.85}As$ shell, a 2 ML InAs QD shell and a 7 nm thick $In_{0.15}Ga_{0.85}As$ shell. The outer 50 nm thick GaAs shell was Si doped by simultaneous supply of Si during shell growth for *n*-type conductivity.[39]

**Transmission electron microscopy** – STEM images were obtained with a JEM-2100F system operating at 200 kV equipped with a HAADF detector for Z-contrast imaging conditions. The plan-view TEM specimen preparation was performed by a two-step procedure adapted from the procedure previously reported for NWs longer than 1 μm.[24,52] First, the NWs were mechanically stabilized on the original substrate using a solution of epoxy Gatan G1 (Gatan) in acetone over the sample. Gatan G1 is very effective in filling out the interwire space and is stable under electron-beam irradiation.[52] In the second step, the NW samples were thinned, from the back-side only, using standard methods of grinding, dimpling, and Argon ion milling until electron transparency was reached (thickness <100 nm). This method provides large thin plan-view sections with a large number of NWs (>200), although it is important to note that the observed features are strongly dependent on the QD area probed by the electron beam, i.e. on how the QDs are "cut" during the thinning process, like in any investigation of QD structures using TEM (cf. planar layers).

Estimations of the In distribution profile are based on the analysis of the HAADF STEM intensity following the procedure in Ref. [18]. The noise of the annular dark-field detector was determined by averaging the intensity $I_{vac}$ of the vacuum region around the NW, imaged under identical conditions as the NW. The noise $I_{vac}$ was then substracted from all measured intensities including at the NW position. HAADF intensity profiles from the noise-corrected micrographs were determined by integrating across areas with about 10 nm width perpendicular to the QW. The determination of the composition using HAADF is based on the dependence of the contrast R, defined as the HAADF intensity of (In,Ga)As, $I_{InGaAs}$, divided by that of GaAs, $I_{GaAs}$ on the In content, x. Thickness variations of the TEM specimen led to strongly varying HAADF intensity distributions in the GaAs reference region which hinders the unambiguous determination of $I_{GaAs}$ and do not arise from a change in composition.[24] In order to compensate for variations in specimen thickness, the GaAs intensity profile was fitted by a polynomial function of 5$^{th}$ order, to estimate GaAs intensity at every single position in the area scan. The measured, noise-corrected intensities were then divided by thickness-corrected $I_{GaAs}$ yielding the measured contrast R, defined as $R = 1 + \frac{x(Z_{In}^{\gamma} - Z_{Ga}^{\gamma})}{(Z_{Ga}^{\gamma} + Z_{As}^{\gamma})}$ where $Z_i$ (i = Ga, As, In) is the atomic number and the factor γ is determined from the analysis of reference planar layers of known composition. In the present case, the In content is estimated assuming γ = 1.7. Indium contents assuming γ = 1.8 are about 5% lower than those obtained using γ = 1.7. These values are within the experimental error of the technique, which is about 15%. Indium contents are obtained with an accuracy of ±2.5%.

**Cathodoluminescence** – CL measurements were performed using a Gatan MonoCL4 system mounted to a Zeiss Ultra55 field-emission scanning electron microscope equipped with a He-cooled sample stage operated at 10 K. The luminescence is collected using a parabolic mirror, passed through a grating monochromator and detected using a liquid $N_2$-cooled, infrared-optimized photomultiplier (InP/(In,Ga)As photocathode). Hyperspectral CL linescans were acquired using a Si CCD with a cut-off at about 1060 nm.

**Photoluminescence** – PL measurements were performed in a continuous-flow He cryostat using either a Ti:Sapphire or He:Ne laser for excitation focused onto the sample by either a 10× or 50× microscope objective. The PL signal was collected by the same microscope objective, dispersed with a monochromator equipped with a 750 lines/mm grating and detected with a liquid $N_2$-cooled (In,Ga)As detector array with a cut-off at ≈1.6 µm.

**LED processing** – LEDs were processed from a DWELL NW sample grown on a *p*-type Si wafer following our previous work.[37] The *p*-type GaAs:Be core of our LED structure provides a direct electrical contact to the substrate. NWs were embedded in benzocyclobutene (BCB) for electrical isolation. The BCB layer was etched down to approximately 1.5 µm by reactive ion etching (RIE) ($CF_4/O_2$) and the NW tips were contacted with sputtered indium tin oxide (ITO). Devices of different size were patterned in the ITO layer by optical lithography and inductively coupled plasma reactive ion etching (ICP-RIE) ($BCl_3/Cl_2$). Back-side Ni/Au and top cross-hair Ti/Au contacts were deposited to improve the electrical contact. Finally, processed samples were mounted and wire bonded in chip carriers.

**Electroluminescence** – LEDs were driven under continuous bias and the EL signal was collected by a lens, dispersed with a monochromator equipped with a 150 lines/mm grating, and detected by a liquid $N_2$-cooled (In,Ga)As array detector with cutoff at about 1.6 µm. The current-voltage characteristics of the investigated devices was recorded using a semiconductor analyzer.

**Acknowledgments**

The authors acknowledge funding from the German Federal Ministry of Education and Research BMBF in the framework of project MILAS. P.C. acknowledges funding from the Fonds National Suisse de la Recherche Scientifique through Project No. 161032. R.B.L. acknowledges funding from the Alexander von Humboldt Foundation. The authors are grateful to M. Höricke, C. Herrmann and C. Stemmler for MBE maintenance, to A.-K. Bluhm for SEM images, to O. Krüger and M. Matalla (Ferdinand-Braun-Institut Berlin) for e-beam lithography, to M. Matzeck and S. Krauß for TEM sample preparation and to S. Rauwerdink and A. Riedel for assistance in device processing. The authors also would like to thank H. Küpers for helpful discussions and advice on patterned substrate preparation and growth. Also, the authors are grateful to W. Prost for helpful discussions on the I-V characteristics of the NW LEDs and to E. Zallo for a critical reading of the manuscript.

(22)  Joyce, B. A.; Vvedensky, D. D. Self-Organized Growth on GaAs Surfaces. *Mater. Sci. Eng. R Reports* **2004**, *46* (6), 127–176.

(23)  Grandal, J.; Wu, M.; Kong, X.; Hanke, M.; Dimakis, E.; Geelhaar, L.; Riechert, H.; Trampert, A. Plan-View Transmission Electron Microscopy Investigation of GaAs/(In,Ga)As Core-Shell Nanowires. *Appl. Phys. Lett.* **2014**, *105* (12), 121602.

(24)  Nicolai, L. Interface Structure and Strain Relaxation of III-V Core-Shell Nanowire Heterostructures Studied by Transmission Electron Microscopy, Master Thesis, Humboldt-Universität zu Berlin, 2016.

(25)  Krause, T.; Hanke, M.; Nicolai, L.; Cheng, Z.; Niehle, M.; Trampert, A.; Kahnt, M.; Falkenberg, G.; Schroer, C. G.; Hartmann, J.; Zhou, H.; Wehmann, H.-H.; Waag, A.. Structure and Composition of Isolated Core-Shell (In,Ga) N/GaN Rods Based on Nanofocus X-Ray Diffraction and Scanning Transmission Electron Microscopy. *Phys. Rev. Appl.* **2017**, *7* (2), 024033.

(26)  Corfdir, P.; Lewis, R. B.; Geelhaar, L.; Brandt, O. Fine Structure of Excitons in InAs Quantum Dots on GaAs(110) Planar Layers and Nanowire Facets. *Phys. Rev. B* **2017**, *96* (4), 045435.

(27)  Lewis, R. B.; Trampert, A.; Luna, E.; Herranz, J.; Pfüller, C.; Geelhaar, L. Bismuth-Surfactant-Induced Growth and Structure of InAs/GaAs(110) Quantum Dots. *Semicond. Sci. Technol.* **2019**, *34* (10), 105016.

(28)  Inoue, T.; Kita, T.; Wada, O.; Konno, M.; Yaguchi, T.; Kamino, T. Electron Tomography of Embedded Semiconductor Quantum Dot. *Appl. Phys. Lett.* **2008**, *92* (3), 031902.

(29)  Dimakis, E.; Jahn, U.; Ramsteiner, M.; Tahraoui, A.; Grandal, J.; Kong, X.; Marquardt, O.; Trampert, A.; Riechert, H.; Geelhaar, L. Coaxial Multishell (In,Ga)As/GaAs Nanowires for Near-Infrared Emission on Si Substrates. *Nano Lett.* **2014**, *14* (5), 2604–2609.

(30)  Küpers, H.; Corfdir, P.; Lewis, R. B.; Flissikowski, T.; Tahraoui, A.; Grahn, H. T.; Brandt, O.; Geelhaar, L. Impact of Outer Shell Structure and Localization Effects on Charge Carrier Dynamics in GaAs/(In,Ga)As Nanowire Core–Shell Quantum Wells. *Phys. Status Solidi - Rapid Res. Lett.* **2019**, *13* (5), 1800527.

(31)  Lähnemann, J.; Hill, M. O.; Herranz, J.; Marquardt, O.; Gao, G.; Al Hassan, A.; Davtyan, A.; Hruszkewycz, S. O.; Holt, M. V.; Huang, C.; Calvo-Almazán, I.; Jahn, U.; Pietsch, U.; Lauhon, L. J.; Geelhaar, L. Correlated Nanoscale Analysis of the Emission from Wurtzite versus Zincblende (In,Ga)As/GaAs Nanowire Core-Shell Quantum Wells. *Nano Lett.* **2019**, *19*, 4448–4457.

(32)  Ravindra, N. M.; Srivastava, V. K. Temperature Dependence of the Energy Gap in Semiconductors. *J. Phys. Chem. Solids* **1979**, *40* (10), 791–793.

(33)  Seravalli, L.; Frigeri, P.; Minelli, M.; Allegri, P.; Avanzini, V.; Franchi, S. Quantum Dot Strain Engineering for Light Emission at 1.3, 1.4 and 1.5 μm. *Appl. Phys. Lett.* **2005**, *87* (6), 63101.

(34)  Popescu, D. P.; Eliseev, P. G.; Stintz, A.; Malloy, K. J. Temperature Dependence of the Photoluminescence Emission from InAs Quantum Dots in a Strained Ga 0.85 In 0.15 As Quantum Well. *Semicond. Sci. Technol* **2004**, *19* (04), 33–38.

(35)  Chen, R.; Liu, H. Y.; Sun, H. D. Electronic Energy Levels and Carrier Dynamics in InAs/InGaAs Dots-in-a-Well Structure Investigated by Optical Spectroscopy. *J. Appl. Phys.* **2010**, *107* (1), 013513.
16

TOC graphic:

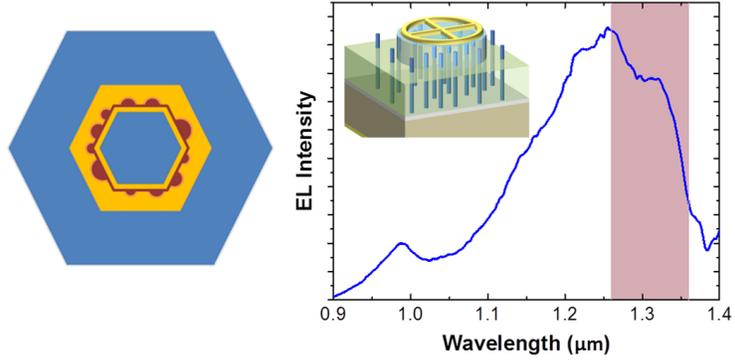